\documentclass[letterpaper, 10 pt, conference]{ieeeconf}  

\IEEEoverridecommandlockouts                              

\usepackage{graphics} 
\usepackage{amsmath} 
\usepackage{graphicx}
\usepackage{subfig}
\usepackage{multirow}
\usepackage{color}
\usepackage{soul}
\definecolor{cyan}{rgb}{0.0, 1.0, 1.0}

\usepackage{bm}

\title{\LARGE \bf
Jam-absorption driving with data assimilation*
}

\author{Siyu Li$^{1}$$^{\dag}$, Ryosuke Nishi$^{2}$, Daichi Yanagisawa$^{3}$, and Katsuhiro Nishinari$^{3}$
\thanks{*This work was supported by JSPS KAKENHI Grant Number JP21H01570, JP21H01352 and JP23K04287, and China Scholarship Council (No. 202108050153).}
\thanks{$^{1}$Siyu Li is with the Department of Advanced Interdisciplinary Studies, School of Engineering,
        The University of Tokyo, Tokyo 153-8904, Japan
        {\tt\small li-siyu@g.ecc.u-tokyo.ac.jp}}%
\thanks{$^{2}$Ryosuke Nishi is with the Department of Mechanical and Physical Engineering, Faculty of Engineering,
        Tottori University, Tottori 680-8552, Japan
        {\tt\small nishi@tottori-u.ac.jp}}%
\thanks{$^{3}$Daichi Yanagisawa and Katsuhiro Nishinari are with the Department of Aeronautics and Astronautics, School of Engineering,
        The University of Tokyo, Tokyo 113-8656, Japan
        {\tt\small tDaichi@mail.ecc.u-tokyo.ac.jp, tknishi@mail.ecc.u-tokyo.ac.jp}}%
\thanks{$^{\dag}$Correspondence:
        {\tt\small li-siyu@g.ecc.u-tokyo.ac.jp}}
}

\begin{document}

\maketitle
\thispagestyle{empty}
\pagestyle{empty}

\begin{abstract}

This paper introduces a data assimilation (DA) framework based on the extended Kalman filter--cell transmission model, designed to assist jam-absorption driving (JAD) operation to alleviate sag traffic congestion. To ascertain and demonstrate the effectiveness of the DA framework for JAD operation, in this paper, we initially investigated its impact on the motion and control performance of a single absorbing vehicle. Numerical results show that the DA framework effectively mitigated underestimated or overestimated control failures of JAD caused by misestimation of key parameters (e.g., free flow speed and critical density) of the traffic flow fundamental diagram. The findings suggest that the proposed DA framework can reduce control failures and prevent significant declines and deteriorations in JAD performance caused by changes in traffic characteristics, e.g., weather conditions or traffic composition.

\end{abstract}

\section{INTRODUCTION}\label{SEC:intro}

\begin{figure*}[t]
    \centering
    \includegraphics[scale=0.195]{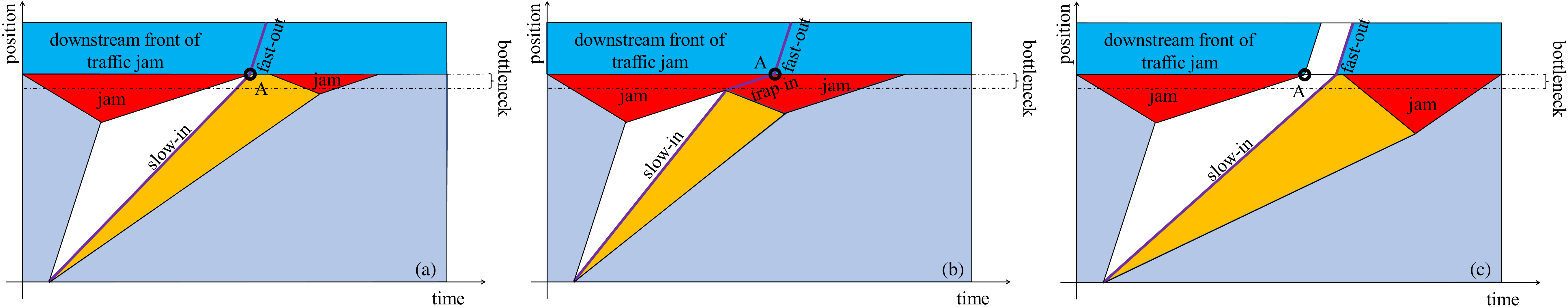}
    \caption{Three cases of JAD: (a) Proper control, (b) Underestimated control, and (c) Overestimated control. Point A denotes an absorbing end point. The white parts imply vacant time space (i.e., no traffic). The purple lines are the trajectories of the absorbing vehicle (AbV).}
    \label{fig:1}
\end{figure*}

Typically, high-demand traffic triggers traffic jams during rush hours. To mitigate freeway traffic jams, variable speed limits using variable speed signs to regulate traffic flows have been developed and applied as one of the most popular control measures \cite{smulders1990control,carlson2011local}. In recent years, with the rapid development of connected and/or automated vehicle (CV/CAV) technologies, an increasing number of researchers have attempted to employ CAVs for alleviating traffic congestion through longitudinal control (LC, i.e., adjusting longitudinal acceleration/deceleration behaviors of a vehicle) \cite{Wang2020LC}. Researchers have proposed and developed the noted adaptive cruise control (ACC); however, generally, it requires at least 20$\sim$30$\%$ market penetration rate (MPR) of ACC-equipped CAVs to have an effect \cite{kesting2008adaptive,goni2019using,kimcacc2022}. Thankfully, jam-absorption driving (JAD) executes straightforward “slow-in” and “fast-out” motions with just a single absorbing vehicle (AbV) to prevent upstream traffic from entering downstream traffic jams, thereby preventing their propagation until they dissipate (see Fig. \ref{fig:1}(a)) \cite{nishi2013theory,taniguchi2015jam,He2017JAD}. Consequently, JAD may function as a substitute for ACC under low CAV MPR environments \cite{li2024jadsystem}.

However, an important factor that significantly affects the JAD performance is the estimation accuracy of the absorbing end point (i.e., point A in Fig. \ref{fig:1}). For instance, He et al. \cite{He2017JAD} demonstrated that either underestimation or overestimation of the absorbing end point results in adverse effects on the control performance. This is because the traffic jam may not be effectively cleared or may even cause further deterioration in the traffic (Fig. \ref{fig:1}(b) and (c)). In addition, the effects of various control speeds in the case of underestimation were analyzed in \cite{han2017variable}. Some studies have attempted to address this issue by dynamically adjusting prediction results based on real-time traffic information \cite{ghiasi2019mixed} or deep reinforcement learning \cite{wang2022trajectory}. However, to the best of our knowledge, none of these studies considered changes in the traffic characteristic \cite{smaragdis2004flow,zhou2022supervised} or errors in traffic models and measurements \cite{wang2005real} in real traffic environments. By ignoring such conditions, these models may still undermine the estimation accuracy of the absorbing end point.

To overcome these limitations of JAD, this paper presents a data assimilation (DA) framework worked by the extended Kalman filter--cell transmission model (EKF-CTM) \cite{zhou2022supervised,tampere2007extended} to reduce the underestimated or overestimated control failures. While numerous studies have employed DA for improving traffic state estimation \cite{wang2005real,tampere2007extended,nantes2016real}, none have integrated it into JAD to improve absorbing end point estimation and control performance. Therefore, this simulation-based study first focuses on investigating the effectiveness of the DA framework for JAD operation. We initially investigated how it affects the motion and control performance of a single AbV against a traffic jam at a sag, a typical freeway bottleneck. The simulation employed a modified intelligent driver model plus (IDM+) \cite{goni2016modeling} to generate virtual true trajectories, followed by using EKF-CTM \cite{zhou2022supervised} to estimate traffic states for performing JAD.

The remainder of this paper is organized as follows. Section \ref{SEC:gro_tru} describes the adopted traffic model for generating virtual true trajectories. Section \ref{SEC:ecfram} presents the EKF-CTM DA framework and how it was embedded into the JAD model. Section \ref{SEC:simu} investigates the effects of the DA on a single AbV. The conclusions are summarized in Section \ref{SEC:concl}.

\section{VIRTUAL TRUE TRAJECTORY GENERATION}\label{SEC:gro_tru}

\subsection{A synthetic single-lane freeway section with a sag}

\begin{figure}[t]
    \centering
    \includegraphics[scale=0.25]{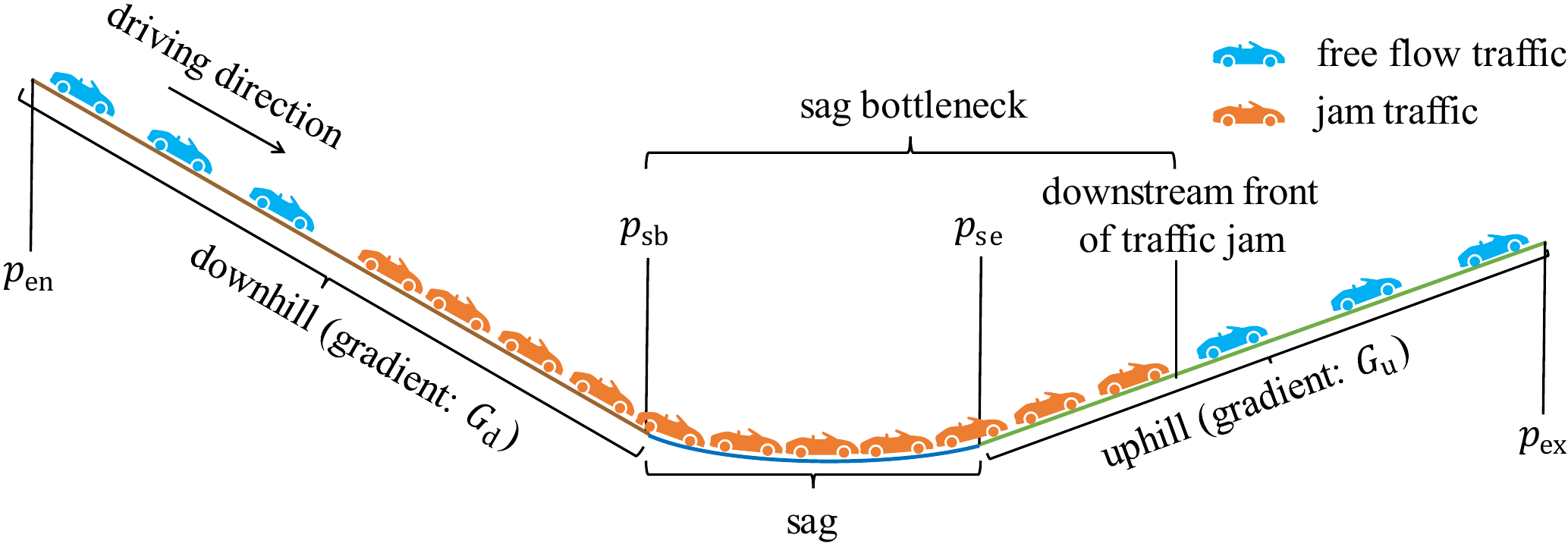}
    \caption{A hypothetical freeway section with a sag.}
    \label{fig:2}
\end{figure}

A sag is a road section where the gradient changes from downhill to uphill. Traffic jams at sags are primarily caused by insufficient acceleration and increased reaction times \cite{yoshizawa2012analysis}. In any given year, sag and uphill sections cause the most traffic jams in Japan \cite{hatakenaka2006development}. Therefore, although this work considered traffic jams generated by a sag, it has not lost its universality and practicality. The geometry of the freeway section is formulated as
\begin{equation}
G(p)=\begin{cases}
\quad \quad \quad G_\text{d} \quad \quad \quad \quad \quad \quad \text{if} \enspace p \le p_\text{sb} \text{,} \\
G_\text{d}+\frac{G_\text{u}-G_\text{d}}{p_\text{se}-p_\text{sb}}\left(p-p_\text{sb}\right) \quad \text{if} \enspace p_\text{sb} < p \leq p_\text{se} \text{,} \\
\quad \quad \quad G_\text{u} \quad \quad \quad \quad \quad \quad \text{if} \enspace p > p_\text{se} \text{,}
\end{cases}
\label{eq1}
\end{equation}

\noindent where $G_\text{d}$ and $G_\text{u}$ are the gradients of the downhill and uphill sections, respectively, $p_\text{sb}$ and $p_\text{se}$ are the positions of the beginning and end of the sag, respectively, as shown in Fig. \ref{fig:2}. The entrance and exit positions of the freeway section are denoted by $p_\text{en}$ and $p_\text{ex}$, respectively. In this work, while we only focused on a single lane to explore the impact of the DA framework, it lays the groundwork for extending the model to address multilane conditions.

\subsection{Car-following model for generating true trajectories}

In this work, a modified IDM+ was used to generate virtual true trajectories. Subsequently, we employed EKF-CTM to estimate the traffic state for performing JAD.

The IDM+ and the slope gradient effect \cite{goni2016modeling} are adopted to model traffic jams triggered by a sag. Specifically, the dynamic acceleration of vehicle $n$ at time $t$ in every time step $\Delta t_\text{vt}$ is formulated as:
\begin{equation}
a_n(t)=a_{\text{des},n}(t)+a_{\text{G},n}(t)\text{.}
\label{eq2}
\end{equation}

The first term $a_{\text{des},n}(t)$ is the desired acceleration determined by the IDM+ \cite{schakel2010effects}, which is a modified version of the original IDM \cite{Treiber2000idm}:
\begin{equation}
\resizebox{\linewidth}{!}{$
a_{\text{des},n}(t)=\alpha\min\left\{1-\left(\frac{v_n(t)}{v_\text{des}}\right)^\delta,1-\left(\frac{s^\ast(v_n(t),\Delta v_n(t))}{s_n(t)}\right)^2\right\}
\text{,}$}
\label{eq3}
\end{equation}

\noindent where $v_n(t)$ is the velocity of vehicle $n$ at time $t$, $\Delta v_n(t)=v_n(t)-v_{n-1}(t)$ is the difference in speed between the leader (vehicle $n-1$) and vehicle $n$ at time $t$, $s_n(t)$ is the bumper-to-bumper space gap, and $s^\ast(v_n,\Delta v_n(t))$ is the desired gap:
\begin{equation}
s^\ast(v_n(t),\Delta v_n(t))=s_0+\max\left\{0,v_n(t)T+\frac{v_n(t)\Delta v_n(t)}{2\sqrt{\alpha\beta}}\right\}
\text{,}
\label{eq4}
\end{equation}

\noindent Parameters $v_\text{des}$, $\delta$, $s_0$, $T$, $\alpha$, $\beta$ are given parameters. Furthermore, the second term $a_{\text{G},n}(t)$ denotes the effect of the road gradient on the acceleration, which is given by \cite{goni2016modeling}
\begin{equation}
a_{\text{G},n}(t)=-\theta \left(G(p_n(t))-G_{\text{c},n}(t)\right)\text{,}
\label{eq5}
\end{equation}

\noindent where $\theta$ is the sensitivity parameter, $G(p_n(t))$ is the road gradient at the position $p_n(t)$, $G_{\text{c},n}(t)$ is the road gradient compensated by vehicle $n$ at time $t$. This implies that the effect of the road gradient on $a_n(t)$ corresponds to the difference between the gradient and the gradient compensation. The compensated road gradient $G_{\text{c},n}(t)$ is updated as
\begin{multline}
G_{\text{c},n}(t)\\=\begin{cases}
G_{\text{c},n}(t-\Delta t_\text{vt})+\lambda \Delta t_\text{vt} \\ \quad \quad \text{if} \enspace G(p_n(t)) > G_{\text{c},n}(t-\Delta t_\text{vt})+\lambda \Delta t_\text{vt} \text{,} \\ \\
G(p_n(t)) \quad \text{else} \text{,}
\end{cases}
\label{eq6}
\end{multline}

\noindent where $\lambda$ represents the maximum gradient compensation rate. Equation \eqref{eq6} implies that: a) vehicle $n$ compensates for an increase in the road gradient with a maximum compensation rate $\lambda$; b) once vehicle $n$ completes its compensation, the compensated road gradient equals the actual road gradient. Furthermore, the model assumes that vehicles can promptly compensate for decreases in gradient. In the simulation, the initial position of vehicle $n$ is located downhill; therefore, the road gradient does not affect the initial acceleration of the vehicles, that is, $G_{\text{c},n}(0)=G(p_n(0))$.

\section{DA FRAMEWORK AND JAD MODEL}\label{SEC:ecfram}

\begin{figure}[t]
    \centering
    \subfloat[JAD with EKF-CTM.]{\label{fig:3sub1}\includegraphics[scale=0.29]{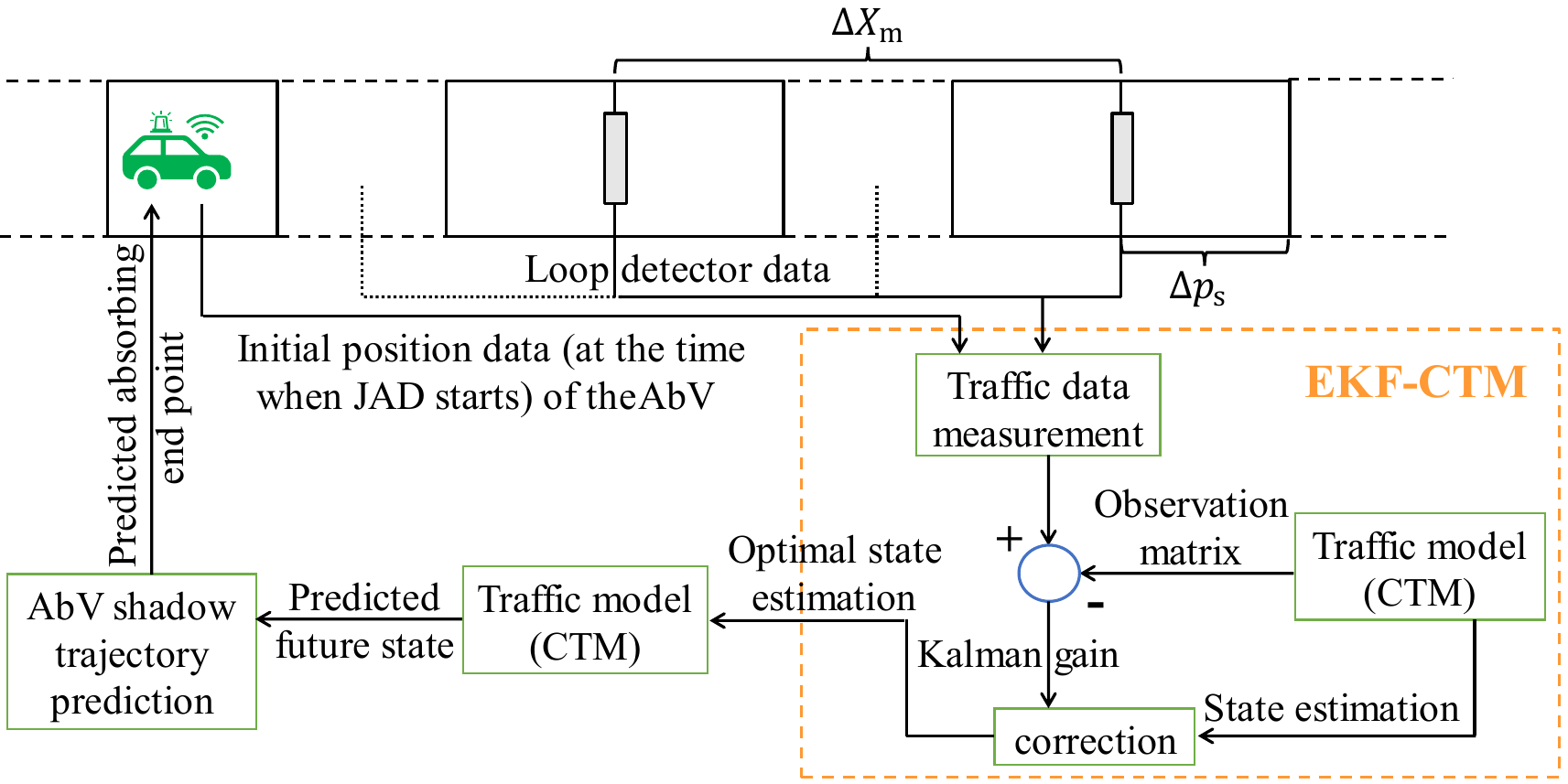}
    }
    \qquad
    \subfloat[Rolling horizon scheme.]{\label{fig:3sub2}\includegraphics[scale=0.35]{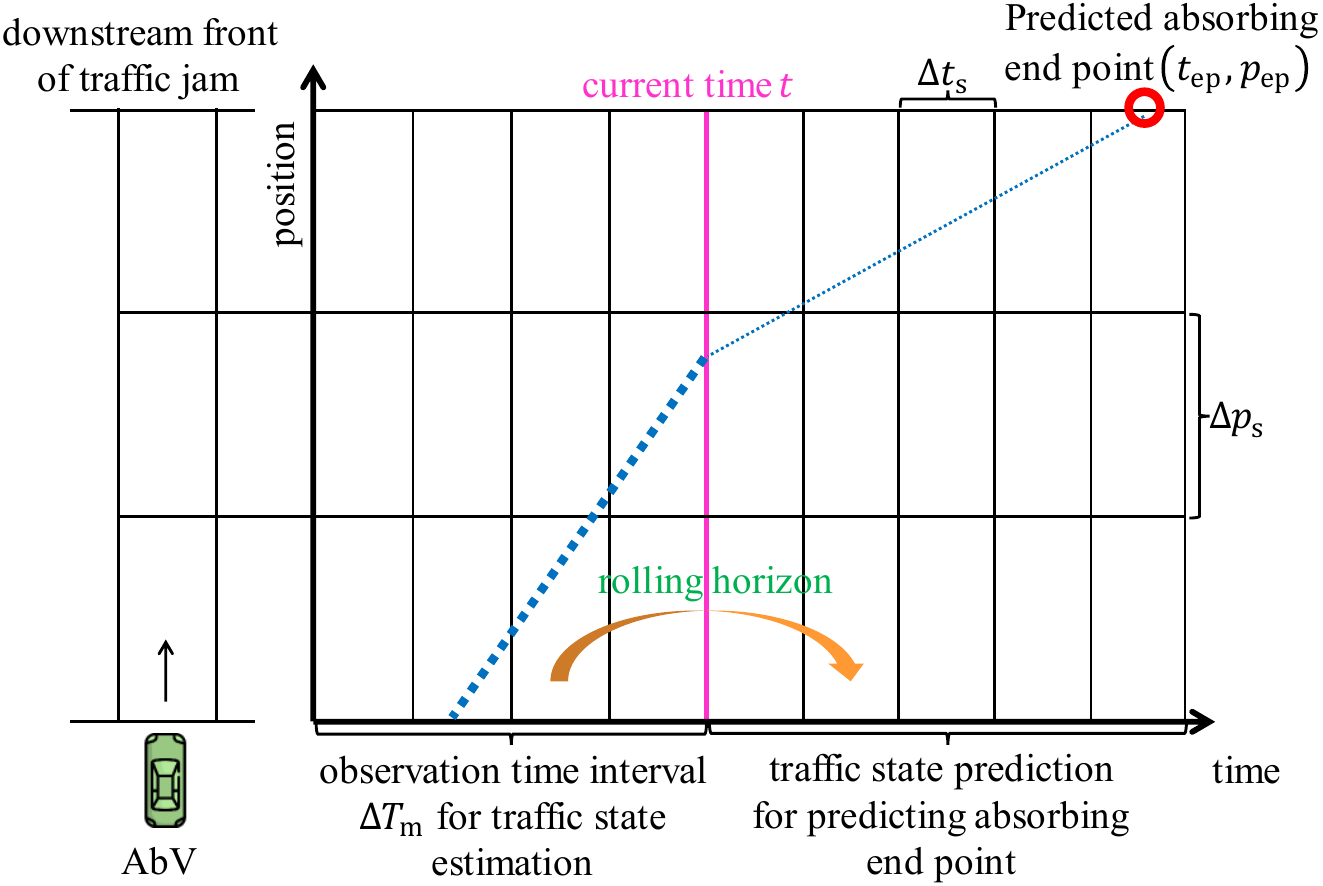}
    }
    \caption{Integrated system framework for JAD with DA.}
    \label{fig:3}
\end{figure}

This section presents the DA framework and how it was embedded into the JAD model. Fig. \ref{fig:3} shows the integrated system framework. The road section was divided into $I$ cells with equal length $\Delta p_\text{s}$ for performing CTM. A rolling horizon scheme was employed for this framework. At the end of each time window (observation period) $\Delta T_\text{m}$, measurement data (flow and time occupancy) collected by loop detectors placed at boundaries of CTM cells with an equal space interval $\Delta X_\text{m}$, are fed to the EKF-CTM for traffic state estimation. Subsequently, the predicted absorbing end point is obtained using the future traffic state calculated from the current time $t$. Finally, the real-time message with the predicted absorbing end point is sent to the absorbing car to perform JAD. Noting that after JAD starts, the absorbing end point is predicted based on the shadow trajectory (the trajectory for the case of no JAD control) of the AbV; therefore, only the initial position (at the time when JAD starts) of the AbV is provided for each JAD.

\subsection{EKF-CTM for traffic state estimation}

The traffic state on a freeway section with total $I$ cells
\begin{equation}
\bm{x}(t)=[\rho_1(t), \rho_2(t), \ldots, \rho_I(t),\bm{\Theta}(t)]^\top
\label{eq7}
\end{equation}

\noindent is modeled over the simulation time step $\Delta t_\text{s}$, includes the densities of all cells $\rho_i(t)$ and key parameters $\bm{\Theta}(t)=[v_\text{fr}(t),\rho_\text{cr,nor}(t),\rho_\text{cr,sag}(t)]$ (i.e., free flow speed, critical densities of normal section and sag-uphill section). Key parameters $\bm{\Theta}(t)$ are jointly estimated with traffic states in real time, as in \cite{zhou2022supervised,nantes2016real}. There are two steps of EKF for each $\Delta t_\text{s}$: forecast and analysis. The following state-space model is used in the forecasting step:
\begin{equation}
\hat{\bm{x}}_\text{-}(t)=f(\hat{\bm{x}}_\text{+}(t-\Delta t_\text{s}))
\label{eq8}\text{,}
\end{equation}
\begin{equation}
\bm{P}_\text{-}(t)=\bm{F}(t-\Delta t_\text{s})\bm{P}_\text{+}(t-\Delta t_\text{s})\bm{F}^\top(t-\Delta t_\text{s})+\bm{Q}(t-\Delta t_\text{s})
\label{eq9}\text{,}
\end{equation}

\noindent where operator $f(\bullet)$ is the traffic model, i.e., sag CTM \cite{daganzo1994cell,jin2018kinematic}, for evolving $\rho_i(t)$. States $\hat{\bm{x}}_\text{-}(t)$ and $\hat{\bm{x}}_\text{+}(t)$ are the estimated mean state vectors before and after the analysis step at each $\Delta t_\text{s}$, respectively. Matrices $\bm{P}_\text{-}(t)$ and $\bm{P}_\text{+}(t)$ are variance-covariance matrices before and after the analysis step at each $\Delta t_\text{s}$, respectively. Matrix $\bm{Q}(t)$ is the system noise matrix. Matrix $\bm{F}(t)$ is the linearization of $f(\bullet)$.

In the analysis step, the mean and covariance matrices of the traffic state are updated using the Kalman filter equations
\begin{equation}
\hat{\bm{x}}_\text{+}(t)=\hat{\bm{x}}_\text{-}(t)+\bm{K}(t)(\bm{y}(t)-h(\hat{\bm{x}}_\text{-}(t)))
\text{,}
\label{eq10}
\end{equation}
\begin{equation}
\bm{P}_\text{+}(t)=\bm{P}_\text{-}(t)-\bm{K}(t)\bm{H}(t)\bm{P}_\text{-}(t)
\text{,}
\label{eq11}
\end{equation}

\noindent where $\bm{y}(t)$ is the measurement matrix,  $\bm{K}(t)$ is the Kalman gain, operator $h(\bullet)$ is the transfer model from the traffic state variable to the measurement state variable, $\bm{H}(t)$ is the linearization of $h(\bullet)$.
\begin{equation}
\bm{K}(t)=\bm{P}_\text{-}(t)\bm{H}^\top(t)(\bm{H}(t)\bm{P}_\text{-}(t)\bm{H}^\top(t)+\bm{R})^{-1}
\text{,}
\label{eq12}
\end{equation}

\noindent where $\bm{R}$ is the measurement noise matrix. Besides, $\hat{\bm{x}}_\text{+}(t)\cong\bm{x}(t)$ is assumed as the true value of the traffic state in the estimation.

\subsection{Prediction of absorbing end point}

When the current time $t$ is the time immediately after the traffic state estimation in each observation period, the temporary position of future shadow trajectory prediction of the AbV is obtained with equilibrium speed $v_{\text{e},i_\text{sht,abv}(t,t)}$ at cell $i_\text{sht,abv}(t,t)$:
\begin{equation}
p_\text{sht,abv}^\text{temp}(t,t+\Delta t_\text{s})=p_\text{sht,abv}(t,t)+v_{\text{e},i_\text{sht,abv}(t,t)}\Delta t_\text{s}
\text{,}
\label{eq13}
\end{equation}

\noindent where the left item of $(t,t)$ denotes the current time, while the right item denotes the future time for prediction. Variable $i_\text{sht,abv}(t,t)$ represents the series number of the cell in which shadow trajectory prediction of the AbV is located at the future time $t$. Position $p_\text{sht,abv}(t,t)$ is the determined position of the shadow trajectory prediction of the AbV at future time $t$. Using $p_\text{sht,abv}^\text{temp}(t, t+\Delta t_\text{s})$, the determined position of the shadow trajectory prediction of the AbV at the next future time instant $t+\Delta t_\text{s}$, $p_\text{sht,abv}(t,t+\Delta t_\text{s})$, is given by
\begin{multline}
p_\text{sht,abv}(t,t+\Delta t_\text{s})\\=\begin{cases}
p_{i_\text{sht,abv}(t,t),\text{db}}+v_{\text{e},i_\text{sht,abv}(t,t)+1}(\Delta t_\text{s}-\Delta t_\text{s}^{\prime}(t,t)) \\ \quad \quad \text{if} \enspace p_\text{sht,abv}^\text{temp}(t,t+\Delta t_\text{s}) > p_{i_\text{sht,abv}(t,t),\text{db}} \text{,} \\ \\
p_\text{sht,abv}^\text{temp}(t,t+\Delta t_\text{s}) \quad \text{else} \text{,}
\end{cases}
\label{eq14}
\end{multline}

\noindent where $p_{i_\text{sht,abv}(t,t),\text{db}}$ is the position of the downstream boundary of cell $i_\text{sht,abv}(t,t)$, $\Delta t_\text{s}^{\prime}(t,t)$ is the estimated time required for the shadow AbV to move from $p_\text{sht,abv}(t,t)$ to $p_{i_\text{sht,abv}(t,t),\text{db}}$, which is given by
\begin{equation}
\Delta t_\text{s}^{\prime}(t,t)=\frac{p_{i_\text{sht,abv}(t,t),\text{db}}-p_\text{sht,abv}(t,t)}{v_{\text{e},i_\text{sht,abv}(t,t)}}
\text{.}
\label{eq15}
\end{equation}

\noindent After $p_\text{sht,abv}(t,t+\Delta t_\text{s})$ is predicted, $p_\text{sht,abv}(t,t+2\Delta t_\text{s})$ is predicted using Equations \eqref{eq13}--\eqref{eq15} in which incrementing future time by $\Delta t_\text{s}$ is replaced by $2\Delta t_\text{s}$. Next, $p_\text{sht,abv}(t,t+3\Delta t_\text{s})$ is predicted using Equations \eqref{eq13}--\eqref{eq15} in which incrementing future time by $2\Delta t_\text{s}$ is replaced by $3\Delta t_\text{s}$. In the same way, $p_\text{sht,abv}(t, t+(\eta+1)\Delta t_\text{s})$ for $\eta \ge 3$ is predicted using Equations \eqref{eq13}--\eqref{eq15} in which incrementing future time by $\eta\Delta t_\text{s}$ is replaced by $(\eta+1)\Delta t_\text{s}$. Thus, positions $p_\text{sht,abv}(t,t+\eta\Delta t_\text{s}) (\eta=1, 2, \ldots)$ are predicted in order.

Because the downstream front of the standing queues at the bottlenecks is approximately fixed \cite{han2017variable,jin2018kinematic,nishi2022system}, the position of the absorbing end point $p_\text{ep}$ can be determined at the downstream front of the traffic jam (near $p_\text{se}$) based on historical observations. Therefore, in this work, $p_\text{ep}$ is assumed to be fixed at a certain location, whereas the time of the absorbing end point predicted at the current time $t$, $t_\text{ep}(t)$, is calculated using the above model. Consequently, $t_\text{ep}(t)$ is dynamically determined using the AbV shadow trajectory prediction, as follows:
\begin{equation}
t_\text{ep}(t)=\acute{t}|_{p_\text{sht,abv}(t,\acute{t})=p_\text{ep}} \text{,} \enspace t < \acute{t} \text{.}
\label{eq16}
\end{equation}

\subsection{JAD}

The acceleration of the AbV is given by:
\begin{equation}
a_\text{abv}(t)=\begin{cases}
a_\text{jad,abv}(t) \quad \text{if} \enspace p_\text{jst} \le p_\text{abv} \le p_\text{jen} \text{,} \\
a_\text{des,abv}(t) \quad \text{else} \text{,}
\end{cases}
\label{eq17}
\end{equation}

\noindent where $a_\text{des,abv}(t)$ is the desired acceleration determined by IDM+, $a_\text{jad,abv}(t)$ is the acceleration control of JAD, $p_\text{abv}(t)$ is the position of the AbV, $p_\text{jst}$ and $p_\text{jen}$ are the positions of the start and end of JAD, respectively. When $p_\text{jst} \le p_\text{abv}(t) \le p_\text{jen}$, the JAD controller aims to manipulate the AbV drive to the predicted absorbing point to avoid entering the downstream traffic jam (i.e., the “slow-in” motion). The $a_\text{jad,abv}(t)$ at each control time step $\Delta t_\text{c}$ is calculated as
\begin{equation}
a_\text{jad,abv}(t)=\begin{cases}
\min\{a_\text{des,abv}(t), a_\text{jad,abv}^\text{temp}(t)\} \quad \text{if} \enspace a_\text{des,abv}(t) < 0 \text{,} \\
\quad \quad \quad a_\text{jad,abv}^\text{temp}(t) \quad \quad \quad \quad \quad \text{else} \text{,}
\end{cases}
\label{eq18}
\end{equation}

\noindent where $a_\text{des,abv}(t)$ is used for crash avoidance. Acceleration $a_\text{jad,abv}^\text{temp}(t)$ is the temporary acceleration control for the JAD and is calculated as follows:
\begin{equation}
\resizebox{\linewidth}{!}{$
a_\text{jad,abv}^\text{temp}=\begin{cases}
\max\{a_\text{si,abv}(t),\underline{a_\text{si}},-\frac{v_\text{abv}(t)}{\Delta t_\text{c}}\} \quad \text{if} \enspace v_\text{si,abv}(t) < v_\text{abv}(t) \text{,} \\
\min\{a_\text{si,abv}(t),\overline{a_\text{si}},\frac{v_\text{des}-v_\text{abv}(t)}{\Delta t_\text{c}}\} \enspace \text{if} \enspace v_\text{si,abv}(t) > v_\text{abv}(t) \text{,} \\
\quad \quad \quad \quad \quad 0 \quad \quad \quad \quad \quad \quad \quad \text{else} \text{,}
\end{cases}$}
\label{eq19}
\end{equation}

\noindent where $v_\text{des}$ is the desired speed of IDM+, $\underline{a_\text{si}}$ and $\overline{a_\text{si}}$ are the lower and upper bounds of the acceleration rate for “slow-in,” respectively. Accelerations $-\frac{v_\text{abv}(t)}{\Delta t_\text{c}}$ and $\frac{v_\text{des}(t)-v_\text{abv}(t)}{\Delta t_\text{c}}$ are set to prevent the control speed from exceeding the range of [$0$ m/s, $v_\text{des}$]. Velocity $v_\text{abv}(t)$ is the speed of the AbV at time $t$, and $a_\text{si,abv}$ and $v_\text{si,abv}$ are acceleration and speed for “slow-in,” respectively.
\begin{equation}
a_\text{si,abv}(t)=\frac{v_\text{si,abv}(t)-v_\text{abv}(t)}{\Delta t_\text{c}}
\text{,}
\label{eq20}
\end{equation}
\begin{equation}
v_{\text{si,abv}}(t)=\min\left\{v_\text{des},\max\left\{\underline{v_\text{si}},\frac{p_\text{ep}-p_\text{abv}(t)}{t_\text{ep}(t)-t}\right\}\right\}\text{,}
\label{eq21}
\end{equation}

\noindent where $\underline{v_\text{si}}$ is the lower bound of the speed for “slow-in,” and $\frac{p_\text{ep}-p_\text{abv}(t)}{t_\text{ep}(t)-t}$ indicates the gradient (i.e., the speed for “slow-in”) between the predicted absorbing end point and current time-space point of the AbV. Additionally, the “fast-out” motion is handled by $a_\text{des,abv}(t)$ in Equation \eqref{eq17}, when $p_\text{abv}(t) > p_\text{jen}$.

\section{NUMERICAL EXPERIMENTS}\label{SEC:simu}

\subsection{Experiment setup}

\begin{figure}[bt]
    \centering
    \includegraphics[scale=0.28]{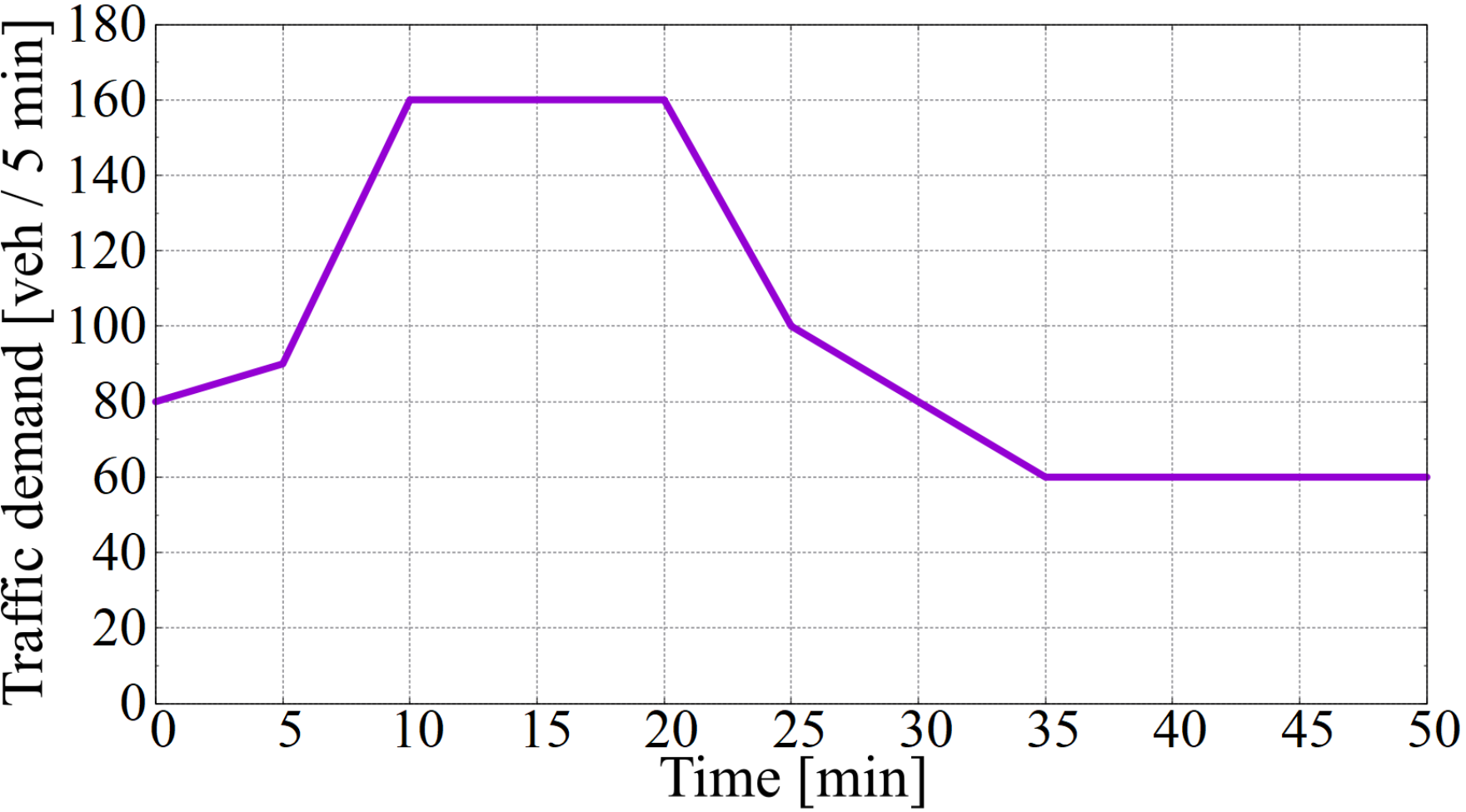}
    \caption{Traffic demand over time.}
    \label{fig:4}
\end{figure}

In the simulation, the gradients of the road section were set as $G_\text{d}=-0.5\%$ and $G_\text{u}=+3.0\%$. Initially, there were no vehicles on freeways. Vehicles $n=1, 2, \ldots$ were generated at $p_\text{en}=0$ m according to Fig. \ref{fig:4}; they then passed $p_\text{sb}=8600$ m and $p_\text{se}=8900$ m. Finally, vehicles were removed from the simulation upon reaching $p_\text{ex}=10000$ m. The parameters related to the modified IDM+ and sag CTM were set according to \cite{goni2019using} and \cite{jin2018kinematic}, respectively. Other parameter settings are listed: $\underline{a_\text{si}}=-1$ m/s$^2$, $\overline{a_\text{si}}=1$ m/s$^2$, $v_\text{des}=27$ m/s, $\underline{v_\text{si}}=5.0$ m/s, $p_\text{jst}=0$ m, $p_\text{jen}=8900$ m, $p_\text{ep}=9200$ m, $\Delta T_\text{m}=60$ s, $\Delta X_\text{m}=500$ m, $\Delta t_\text{vt}=0.01$ s, $\Delta t_\text{s}=3$ s, $\Delta p_\text{s}=100$ m, $\Delta t_\text{c}=0.1$ s.

\begin{figure*}[t]
    \centering
    \subfloat[Balseline: (\romannumeral1) Non-JAD; (\romannumeral2) Proper JAD without DA ($v_\text{fr,ini}=27.0$ m/s, $\rho_\text{cr,nor,ini}=23.0$ veh/km, $\rho_\text{cr,sag,ini}=18.0$ veh/km).]{\label{fig:5sub1}\includegraphics[scale=0.235]{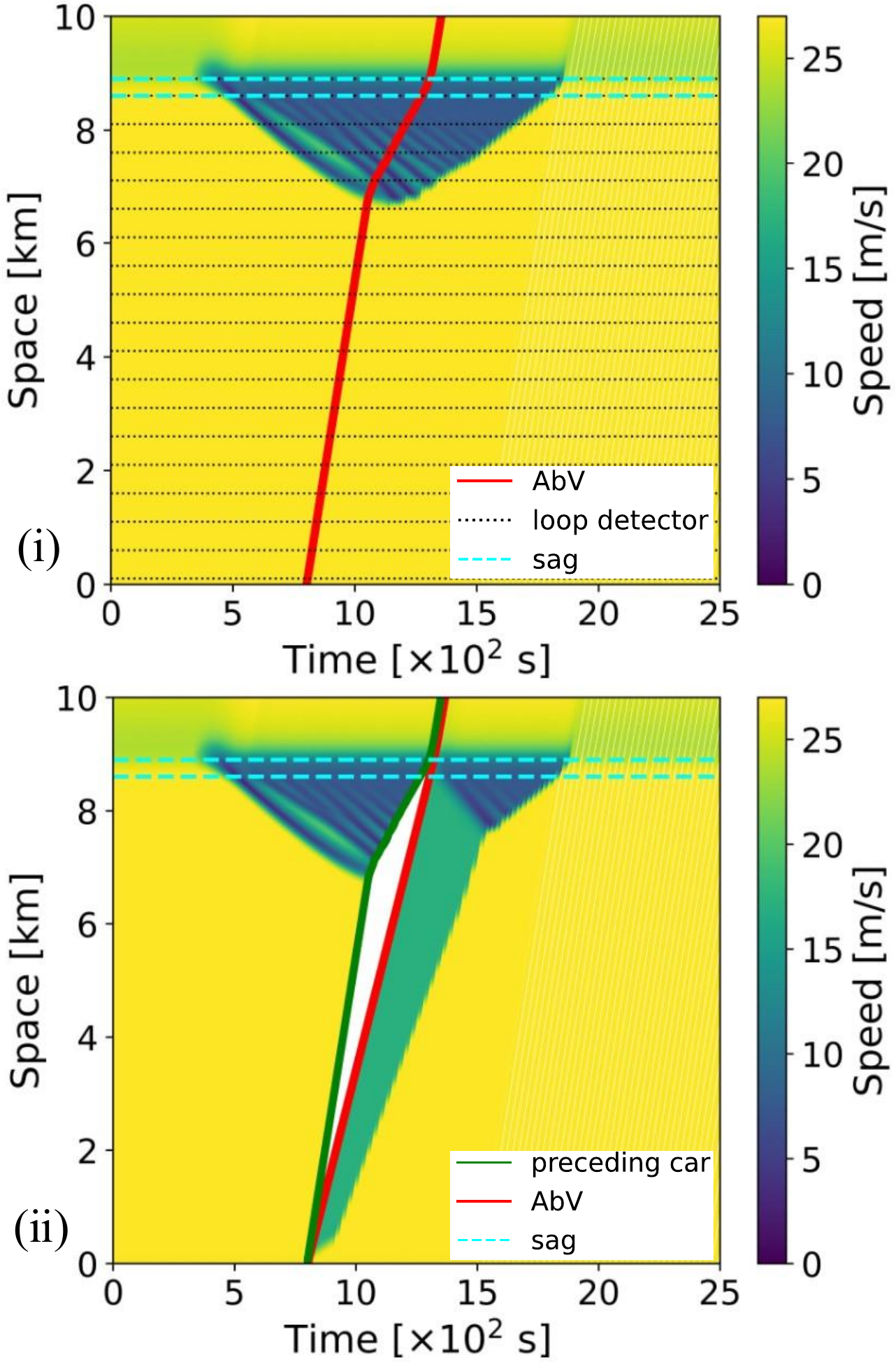}
    }
    \quad\quad\quad
    \subfloat[UE ($v_\text{fr,ini}=30.0$ m/s, $\rho_\text{cr,nor,ini}=23.0$ veh/km, $\rho_\text{cr,sag,ini}=18.0$ veh/km): (\romannumeral1) JAD without DA; (\romannumeral2) JAD with DA.]{\label{fig:5sub2}\includegraphics[scale=0.235]{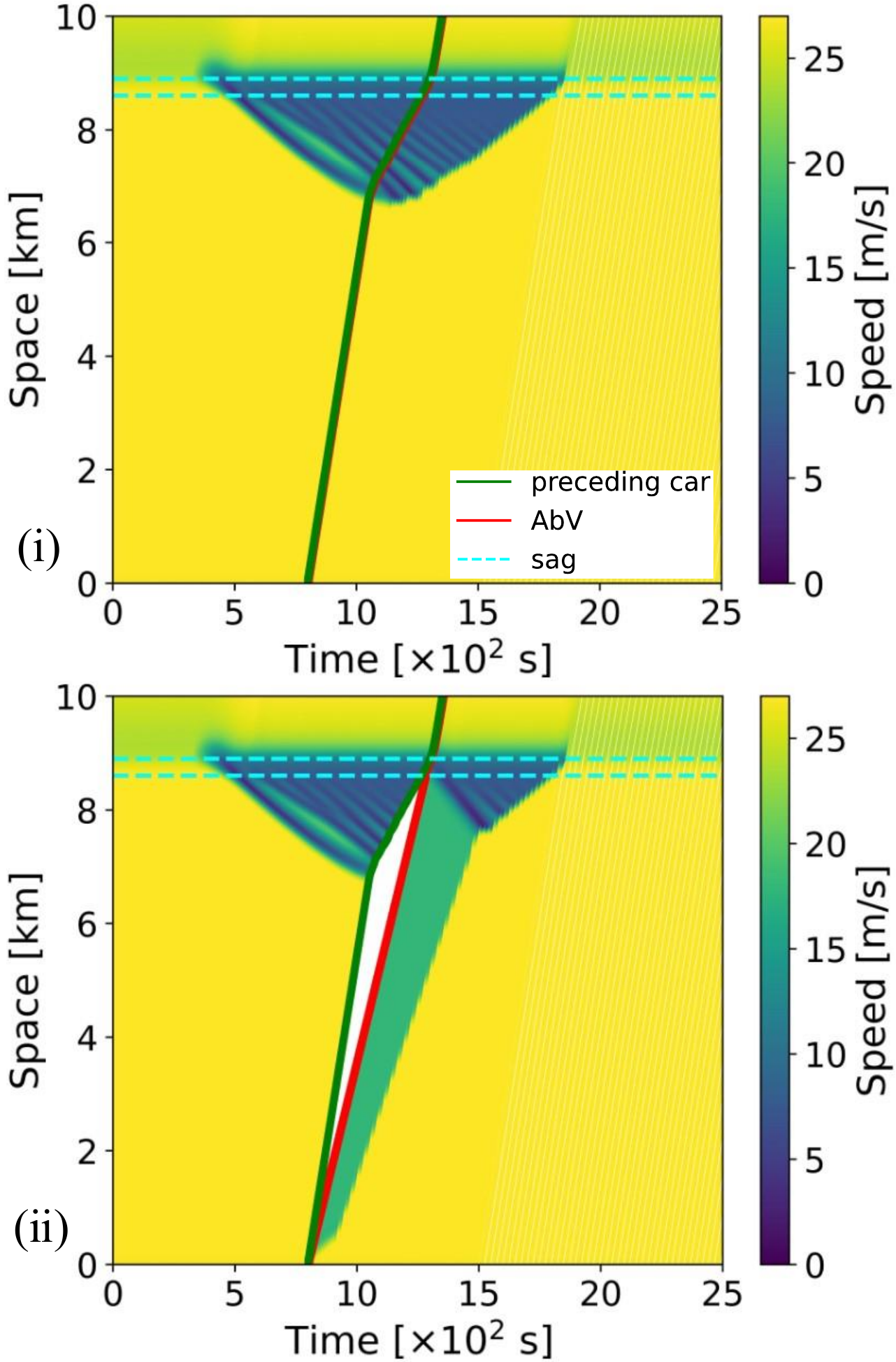}
    }
    \quad\quad\quad
    \subfloat[OE ($v_\text{fr,ini}=24.0$ m/s, $\rho_\text{cr,nor,ini}=26.0$ veh/km, $\rho_\text{cr,sag,ini}=15.0$ veh/km): (\romannumeral1) JAD without DA; (\romannumeral2) JAD with DA.]{\label{fig:5sub3}\includegraphics[scale=0.235]{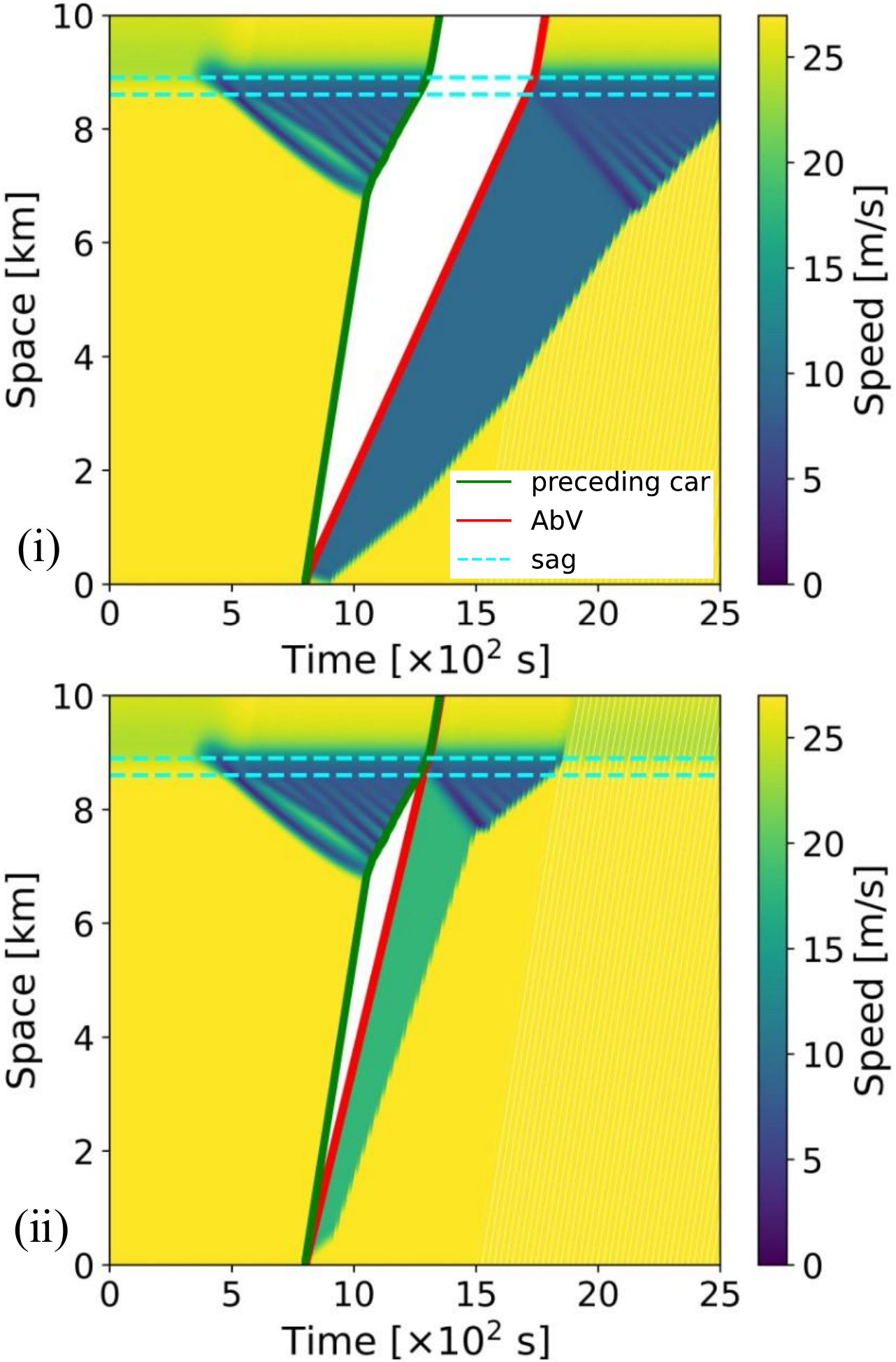}
    }
    \caption{Time-space diagrams (loop detectors are not displayed except for in (a)-\romannumeral1 \enspace for clarity).}
    \label{fig:5}
\end{figure*}

\begin{figure}[!b]
    \centering
    \subfloat[$\Delta ATT$.]{\label{fig:6sub1}\includegraphics[scale=0.38]{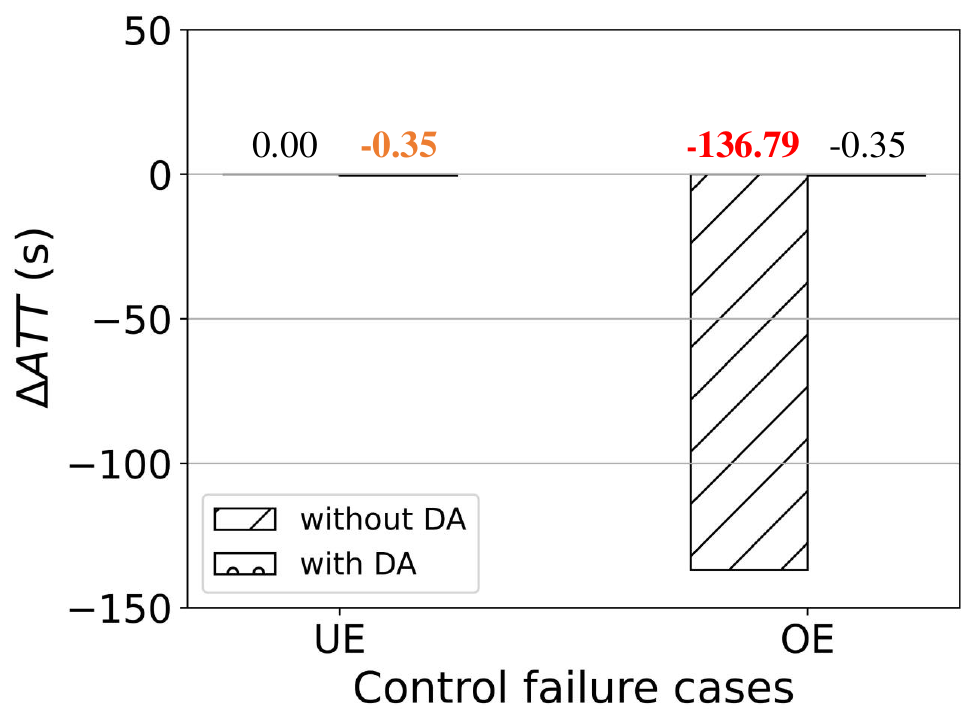}
    }
    \qquad
    \subfloat[$\Delta AFC^\text{Vm}$.]{\label{fig:6sub2}\includegraphics[scale=0.38]{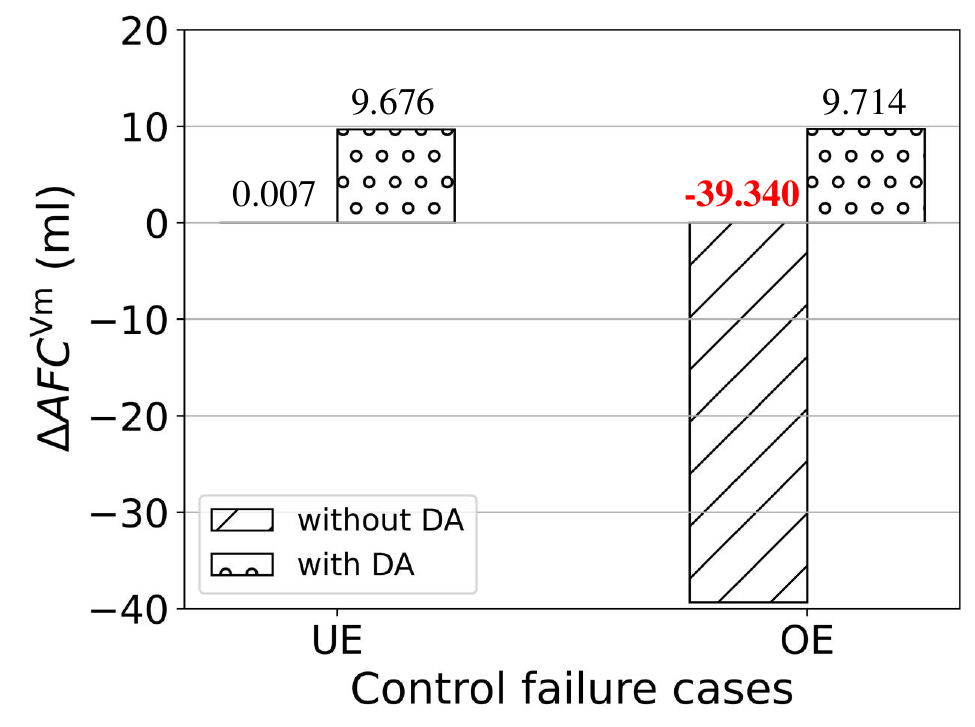}
    }
    \caption{Benefits of without and with DA under different control failure cases.}
    \label{fig:6}
\end{figure}

\subsection{Results}

Figure \ref{fig:5}\subref{fig:5sub1} first presents baseline examples. Without JAD (Fig. \ref{fig:5}\subref{fig:5sub1}-\romannumeral1), standing queues propagate upstream while broking down into stop-and-go waves; traffic jams stop propagate and dispissate when traffic demand decreases. Moreover, according to the AbV trajectory, the lower velocity in the traffic jam delays vehicles traveling through the road section. Fig. \ref{fig:5}\subref{fig:5sub1}-\romannumeral2 \enspace exhibits a proper JAD without DA, with default initial value of key parameters: $v_\text{fr,ini}=27.0$ m/s, $\rho_\text{cr,nor,ini}=23.0$ veh/km, $\rho_\text{cr,sag,ini}=18.0$ veh/km.

In addition, to evaluate the effectiveness of the DA framework, a sensitivity analysis of the various initial values of key parameters was performed. Fig. \ref{fig:5}\subref{fig:5sub2} and \subref{fig:5sub3} shows the time-space diagrams of JAD of the two typical cases of control failure caused by misestimation of these key parameters: underestimation (UE, $v_\text{fr,ini}=30.0$ m/s, $\rho_\text{cr,nor,ini}=23.0$ veh/km, $\rho_\text{cr,sag,ini}=18.0$ veh/km) and overestimation (OE, $v_\text{fr,ini}=24.0$ m/s, $\rho_\text{cr,nor,ini}=26.0$ veh/km, $\rho_\text{cr,sag,ini}=15.0$ veh/km). The figure illustrates that the DA framework corrected the prediction of the absorbing end point, thus achieving proper JAD.

Moreover, indices $\Delta ATT=ATT_\text{nj}-ATT_\text{j}$ and $\Delta AFC^\text{Vm}=AFC_\text{nj}^\text{Vm}-AFC_\text{j}^\text{Vm}$, which indicate travel time reduction per vehicle (s) and fuel consumption reduction per vehicle (ml), were used to compare the benefits between without and with DA. Here, $ATT_\text{nj}$ and $AFC_\text{nj}^\text{Vm}$ are the average travel time per vehicle (s) and average fuel consumption per vehicle (ml) estimated by a typical fuel consumption model (VT-micro) \cite{ahn1998microscopic}, in the case of non-JAD; whereas $ATT_\text{j}$ and $AFC_\text{j}^\text{Vm}$ are the same indices, in the case of JAD. The results are summarized in Fig. \ref{fig:6}. Red value labels represent significant deteriorations in traffic performance (significant deterioration for $\Delta ATT$ was considered as $< -6$ s in this work); orange value labels indicate that the improvements of the with-DA scenario are less than those of the without-DA scenario. The figure indicates that without DA, control performance might deteriorate because of control failures; whereas with DA, control performance generally improved as control failures were reduced.

Usually, on the one hand, UE can maintain or slightly improve traffic performance \cite{piacentini2018traffic}, as shown in Fig. \ref{fig:6}; on the other hand, with DA, the more accurate absorbing end point prediction for JAD would slightly increase the time headway of the AbV when leaving the bottleneck, which results in slightly decrease on $\Delta ATT$. Fortunately, Ghiasi et al. \cite{ghiasi2019mixed} and Nishi and Watanabe \cite{nishi2022system} demonstrated that sequential JAD with multiple AbVs has the potential to suppress this phenomenon, thereby improving $\Delta ATT$.

In short, without DA, the underestimated/overestimated control due to changes in key parameters (changes in traffic characteristics) may lead to significant declines and deteriorations in the control performance; with DA, these adverse effects can be avoided.

\begin{figure}[t]
    \centering
    \includegraphics[scale=0.42]{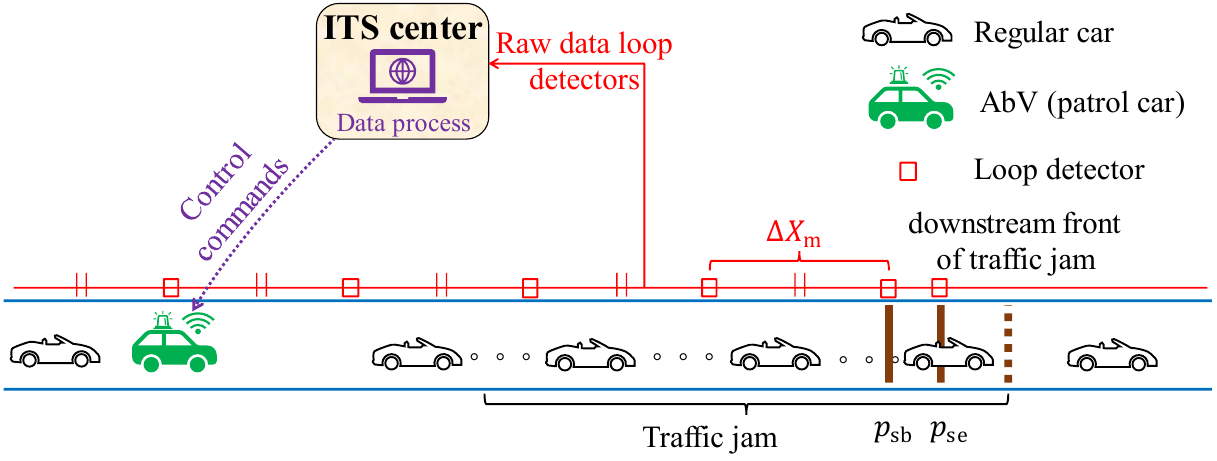}
    \caption{A hypothetical operational system.}
    \label{fig:7}
\end{figure}

\section{CONCLUSIONS}\label{SEC:concl}

In this work, we have proposed an EKF-CTM DA framework to improve JAD control performance in mitigating sag traffic congestion. To ascertain and demonstrate the effectiveness of the DA framework, we have first investigated how it affects the motion and control performance of a single AbV. The numerical results showed that: the DA framework could revise the prediction of absorbing end point, thereby reducing control failures due to misestimation of key parameters (i.e., free-flow speed, critical densities of the normal section and sag-uphill section) of the fundamental diagram. This suggests that the proposed DA framework can sustain effective JAD control when traffic characteristics change, e.g., weather conditions or traffic composition.

Additionally, this work has demonstrated that the DA framework effectively assisted JAD with only measurement data from loop detectors. Fig. \ref{fig:7} shows a hypothetical operational system based on this work. Under the coordination of the Intelligent Transportation System (ITS) center, patrol cars belonging to the freeway management department can be dispatched as AbVs from the upstream on-ramp. Communications among loop detectors, the ITS center, and patrol cars can rely on the existing dedicated communication and transmission systems. These imply that the proposed framework can be applied and implemented in the absence of advanced and expensive equipment that is not easily and rapidly popularized in developing countries and regions in the short run, e.g., roadside units, cameras or lidars, and private CVs/CAVs.

Future work should thoroughly examine the robustness of the DA framework, such as testing its stability under various traffic conditions. Additionally, although the current work only considered single-lane and homogeneous traffic conditions, it lays the groundwork for extending the model to multiclass and multilane traffic scenarios because the effectiveness of the DA framework has been demonstrated; in future work, we would like to extend the proposed framework to more complex scenarios (e.g., multiclass traffic and multiple lanes with lane-changing behaviors) on freeway sections or urban streets.

\addtolength{\textheight}{-12cm}   


\bibliography{ieeeproc}
\bibliographystyle{ieeetr}

\end{document}